# Experimental evidence of forward energy transfer induced by core-edge coupling in multiscale tokamak turbulence

F. Chen[1,7,*], Z.M. Shen[2,7], N. Sun[1,3,7], P. Li[1,*], Z.Y. Qiu[1], P. Liu[4,*], Z. Lin[5], H.T. Chen[6], J.S. Geng[1], Y.X. Li[1,3], P.J. Sun[1], E.Z. Li[1], X.M. Zhong[1], A.D. Liu[2], L. Zhang[1], J. Huang[1], Y.M. Duan[1], T.F. Zhou[1], H.L. Zhao[1], Q. Zang[1], Y.Q. Chu[1], S.X. Wang[1], H.Q. Liu[1], T. Zhang[1], Y.F. Jin[1], B. Lyu[1], C.B. Wu[1], Y.C. Hu[1], L. Ye[1], X.J. Wang[1], X.Z. Gong[1], Y.D. Li[1] and J.G. Li[1,*]

[1] *Institute of Plasma Physics, Hefei Institutes of Physical Science, Chinese Academy of Sciences, Hefei 230031, China*

[2] *Peking University, Beijing 100871, China*

[3] *University of Science and Technology of China, Hefei 230026, China*

[4] *Future Energy Institute, Shanghai Jiao Tong University, Shanghai 201210, China*

[5] *Department of Physics and Astronomy, University of California, Irvine, CA 92697, USA*

[6] *Southwestern Institute of Physics, Chengdu 610041, China*

[7] *These authors contributed equally*

## Abstract

Turbulence in magnetically confined plasmas is inherently quasi-two-dimensional and can exhibit both inverse and forward energy transfer across scales. Experimental evidence for cross-scale energy transfer in the hot plasma core has, however, remained scarce. Understanding such energy transfer is crucial because it governs turbulent transport, which directly impacts confinement quality of fusion plasmas. Here we report experimental evidence consistent with forward energy transfer associated with core-edge coupling in tokamak turbulence. We combined multichannel collective Thomson scattering (CTS) and radially localized Doppler backscattering (DBS) to resolve cross-scale fluctuation dynamics. The results reveal that a quasi-coherent mode (QCM) originating from the edge is also present in the core and resonantly couples with broadband turbulence, producing significant forward energy transfer and characteristic secondary instabilities. These combined measurements indicate that quasi-two-dimensional tokamak turbulence can support a net energy transfer from ion to electron scales through cross-scale wave-wave interactions, contrary to the inverse cascade paradigm of idealized two-dimensional turbulence. Such a plasma regime is attractive for fusion reactors, as it avoids large edge-localized modes (ELMs) that damage first-wall components and maintains high confinement quality even in the presence of high-Z impurities.

---

Many fluid systems in nature and laboratory are constrained to evolve in approximately two dimensions, exhibiting quasi-two-dimensional (quasi-2D) turbulence(*1*). This dimensional constraint fundamentally alters energy transfer. In three dimensions (3D), energy cascades forward from large to small scales, ultimately dissipating at the smallest scales. In two dimensions (2D), by contrast, energy transfers inversely from small to large scales, self-organizing into coherent structures(*2*). Recent studies have shown that quasi-2D systems can support both forward and inverse energy transfer channels simultaneously, enabling seemingly structureless turbulent cascades to generate ordered patterns(*3*). Understanding how energy is transferred across scales is therefore essential for developing a comprehensive framework of turbulence and its transport behaviour.

Tokamak turbulence is a prime example of such quasi-2D systems. Utilizing a toroidal magnetic field configuration to confine a hot plasma, the tokamak approach is one of the most attractive designs for developing nuclear fusion reactors(*4*). Despite this promise, the turbulence in these devices drives anomalous transport and remains a major obstacle to achieving fusion energy(*5*). While the inverse energy transfer channel and its associated zonal flows have been extensively studied as a turbulence suppression mechanism(*6-8*), whether a detectable forward channel also exists in tokamak turbulence has remained an open question. Since forward transfer in 3D turbulence is intimately linked to energy dissipation, the possible existence of such a channel naturally raises the question of whether it correlates with confinement improvement.

Experimentally, energy transfer across scales has been resolved in a variety of fluid systems, from optically accessible electrolyte solutions using visible light imaging(*9*) to low-temperature plasmas using local probes(*10, 11*). Such direct measurements, however, are not feasible in the high-temperature core of fusion plasmas, where temperatures exceed $10^7$ K and no solid probe can survive. Following the landmark work of Surko, Slusher, and Mazzucato(*12-14*), collective Thomson scattering (CTS) using electromagnetic waves as probes offers a non-invasive method for measuring multiscale turbulence in such extreme environments. A central challenge has been to use these scattering measurements to determine the direction of energy transfer. Spectral scaling approaches are complicated by the widely varying decay index across devices and regimes(*15*). Cross-coherence analysis among CTS channels at different wavenumbers also faces a fundamental limitation: in a turbulent cascade, the dominant frequency typically varies with wavenumber, so different channels detect distinct characteristic frequencies. Same-frequency cross-coherence therefore yields negligible coherence between channels(*16*). Consequently, neither method appears capable of resolving the dynamics of energy transfer in the core of magnetically confined fusion plasmas.

Here we report experimental evidence of forward energy transfer in multiscale core turbulence in a superconducting tokamak for the first time. Our observations are enabled by a previously unrecognized turbulent regime in which a quasi-coherent mode (QCM) originating from the edge coexists with broadband turbulence in the core. In this core-edge coupling regime, the QCM nonlinearly interacts with broadband turbulence, generating cross-scale structures that allow the forward energy transfer to be captured by cross-coherence analysis among CTS channels at different wavenumbers. This contrasts with conventional CTS measurements, which have typically detected only broadband fluctuations without such coherent patterns(*17-21*). Moreover, while earlier studies of energy transfer in tokamak edge turbulence did not establish a connection to confinement(*10*), our results show that the forward-transfer regime is associated with improved confinement in long-pulse discharges and, in certain cases, spontaneous plasma oscillations that affect discharge control. These repeatable results demonstrate forward energy transfer from ion to electron scales in tokamak plasma turbulence, highlighting the potential of this process to mitigate turbulent transport and improve confinement.

## Results

### Anomalous confinement improvement with increased radiation loss in long-pulse discharges

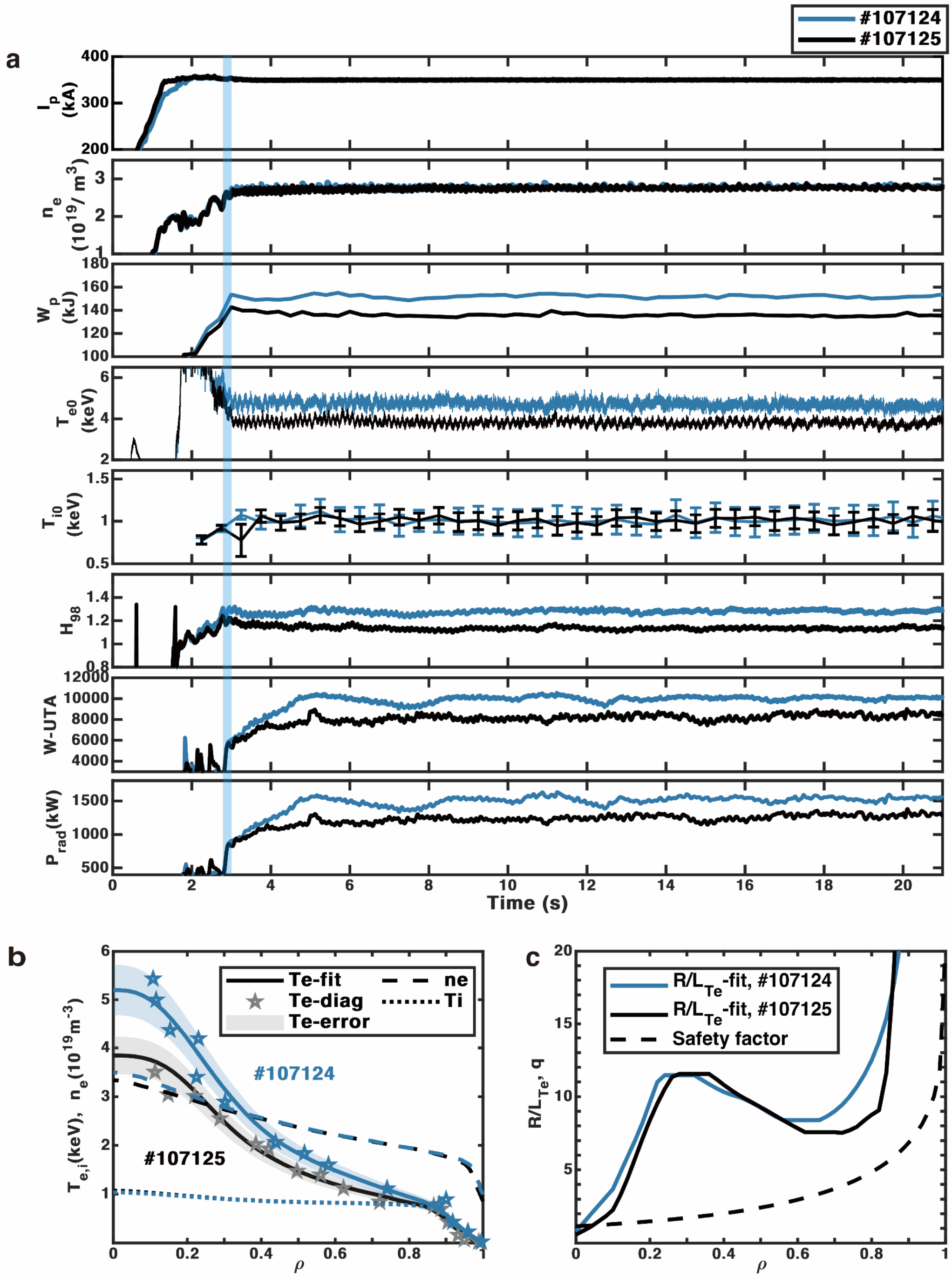

**Fig. 1 | Time traces of long-pulse discharges with identical auxiliary heating power but exhibiting a confinement bifurcation.** Blue colour denotes the improved-confinement discharge (#107124) and black the reference discharge (#107125). **a**, Time evolution of main plasma parameters shows bifurcation of confinement. **b**, Density and temperature profiles. **c**, Electron temperature gradient and the safety factor profiles.

The experiments are conducted on the Experimental Advanced Superconducting Tokamak (EAST)(*22*), a fully superconducting tokamak with full metal walls. Its all-metal, tungsten-based first wall makes EAST directly relevant to ITER(*23*) and future burning plasma devices, where understanding the impact of high-Z impurities on core confinement is essential. This article reports results from two experimental campaigns. In the 2022 campaign, long-pulse discharges were employed to establish the fundamental signatures of forward energy transfer in stationary H-mode conditions. The 2024 campaign extended these observations to short-pulse discharges, revealing how this energy transfer influences discharge evolution and steady-state control, and providing complementary experimental evidence for the core-edge coupling mechanism.

For the long-pulse discharges, an overview of two typical shots is shown in Fig. 1a. Although the operation parameters were closely matched for the two discharges, including the plasma current, equilibrium, density and auxiliary heating scheme (upper panels in Extended Data Fig. 1), a dramatic bifurcation of electron thermal confinement is clearly seen after the injection of ion-cyclotron-range-of-frequency (ICRF) power, as indicated by the vertical light blue rectangle near $t$ = 3 s. This bifurcation is reflected in the stored energy, core electron temperature, and $H_{98}$ factor (a measure of energy confinement quality)(*24*). Discharge #107124, which features improved energy confinement, also showed higher radiated power and stronger high-Z impurity emission than the reference discharge. In particular, the tungsten unresolved transition array (W-UTA) signal, derived from $\mathrm{W}^{27+}-\mathrm{W}^{45+}$ line emissions in the extreme-ultraviolet range(*25*), was enhanced in #107124. While the ion temperature and electron density profiles for both shots are very close, the electron temperature for #107124 is higher than that of #107125 over a wide radial range (Fig. 1b). The largest difference in electron temperature gradient lies inside the pedestal region, at $\rho\sim(0.6,\ 0.8)$, as shown in Fig. 1c. Here, $\rho$ is the normalized poloidal flux coordinate, commonly referred to as the normalized minor radius, and the pedestal region refers to the plasma edge region with steep pressure gradients. Meanwhile, the density feedback required a lower supersonic molecular beam injection (SMBI) rate in #107124 (Extended Data Fig. 1), consistent with reduced outward particle transport. Extended Data Fig. 2 further reveals that, while low-Z impurities such as lithium, carbon, nitrogen and oxygen remain at comparable levels between the two discharges, high-Z impurities including iron, copper, molybdenum and tungsten are systematically elevated in #107124, in line with the observed increase in radiation. Given that high-Z impurities generally degrade confinement, this simultaneous improvement in confinement and impurity accumulation presents a paradox. These observations suggest modifications in turbulent transport, a possibility examined through direct turbulence and energy transfer measurements in the following sections. In this work, we use $k_1, k_2, k_3$ to denote the poloidal wavenumber channels of the CTS diagnostic. Their correspondence to actual $k_\theta$ values differs between the two experimental campaigns; a summary is provided in Table 1. In both campaigns, the subscript increases with wavenumber.

**Table 1 | Correspondence between channel labels and poloidal wavenumbers in the 2022 and 2024 EAST campaigns.** In the 2022 campaign, only two CTS channels were available. An additional channel at 5 $\mathrm{cm}^{-1}$ was installed for the 2024 campaign.

| Campaign | $k_1$ | $k_2$ | $k_3$ | Shot Number |
|---|---|---|---|---|
| 2022 (long-pulse) | 10 $\mathrm{cm}^{-1}$ | 20 $\mathrm{cm}^{-1}$ | - | #107124, #107125 |
| 2024 (short-pulse) | 5 $\mathrm{cm}^{-1}$ | 10 $\mathrm{cm}^{-1}$ | 20 $\mathrm{cm}^{-1}$ | #136748, #136820, #136715 |

## Nonlinear interactions, secondary instabilities and forward energy transfer

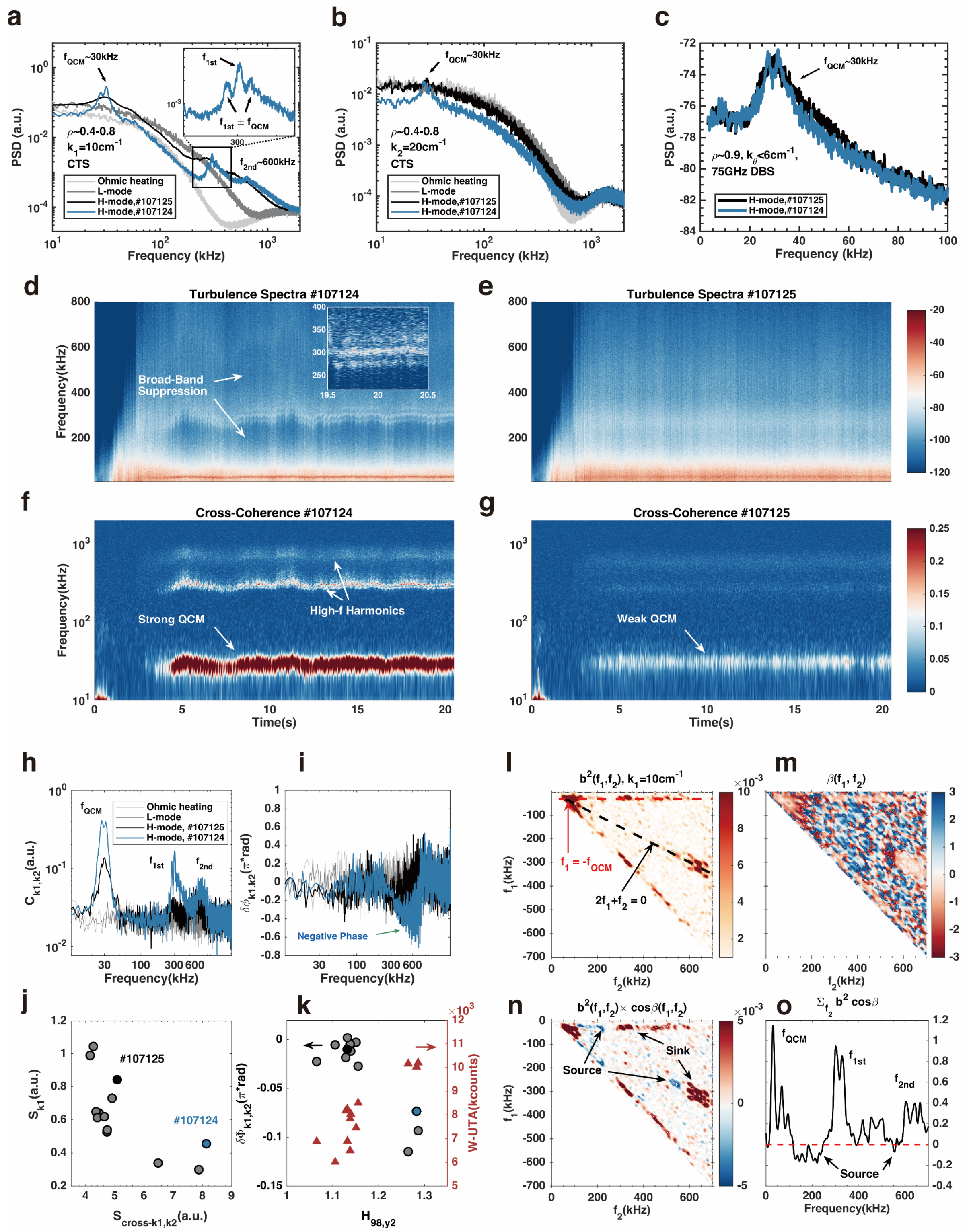

**Fig. 2 | Turbulence characteristics for enhanced and unenhanced confinement regimes. a**, Core turbulence spectra from CTS at $k_1 = 10\ \mathrm{cm}^{-1}$. **b**, Core turbulence spectra from CTS at $k_2 = 20\ \mathrm{cm}^{-1}$. **c**, Edge turbulence spectra from DBS. **d**, **e**, Time evolution of core turbulence spectra from CTS at $k_1 = 10\ \mathrm{cm}^{-1}$ for **d**, #107124 and **e**, #107125. **f**, **g**, Time evolution of cross-coherence between $k_1 = 10\ \mathrm{cm}^{-1}$ and $k_2 = 20\ \mathrm{cm}^{-1}$ for **f**, #107124 and **g**, #107125. **h**, Cross-coherence spectra $C_{k1,k2}$. **i**, Cross-Phase $\delta\phi_{k1,k2}$. The negative phase in **i** indicates forward energy transfer from low-$k$ to high-$k$. The noise level involved in calculation is about $1/\mathrm{N}{\sim}0.02$ for $C_{k1,k2}$ and $0.02\pi$ for $\delta\phi_{k1,k2}$, where N is the ensemble number. **j**, Scatter plot of broadband turbulence intensity $S_{k1}$ versus cross-power $S_{\mathrm{cross}-k1,k2}$. **k**, Averaged cross-coherence phase $\delta\Phi_{k1,k2}$ (circles, left y-axis) and W-UTA signal (triangles, right y-axis) versus $H_{98,y2}$. The data are

collected from 13 H-mode shots with the same auxiliary heating power. **l**, **m**, **n**, **o**, Auto-bicoherence analysis for $k_1 = 10\ \mathrm{cm}^{-1}$at *t*=20 s, #107124. **l**, Auto-bicoherence spectra $b^2(f_1, f_2)$. **m**, Auto-biphase spectra $\beta(f_1, f_2)$. **n**, Energy transfer function $b^2(f_1, f_2) \times \cos\beta(f_1, f_2)$. **o**, Integrated energy transfer function $\sum_{f_2} b^2 \cos\beta$.

We first examine the turbulent fluctuations that distinguish the two confinement regimes. Fig. 2 presents turbulence characteristics across four representative phases: an Ohmic phase, an L-mode phase, the reference H-mode (#107125), and the improved H-mode (#107124). In the core region, the improved-confinement discharge #107124 at $k_1 = 10\ \mathrm{cm}^{-1}$ exhibits a QCM at $f_{\mathrm{QCM}}$~30 kHz, reduced broadband turbulence ($30\ \mathrm{kHz} < f < 300\ \mathrm{kHz},\ 300\ \mathrm{kHz} < f < 600\ \mathrm{kHz}$), and high-frequency harmonics at $f_{\mathrm{1st}}$~300 kHz and $f_{\mathrm{2nd}}$~600 kHz (Fig. 2a, blue line). The unenhanced H-mode shows qualitatively similar but weaker features (Fig. 2a, black line). At $k_2 = 20\ \mathrm{cm}^{-1}$, the high-frequency peaks disappear in #107124, and only the QCM and reduced broadband turbulence remain (Fig. 2b). A magnified view of the spectrum in Fig. 2a reveals clear evidence of secondary instabilities near the first high-frequency harmonic, which strongly resemble features of modulational instability, with sidebands satisfying $f_{\mathrm{sidebands}} = f_{\mathrm{1st}} \pm f_{\mathrm{QCM}}$, a hallmark of nonlinear wave-wave interactions between the QCM and the high-frequency peak. At the edge, only QCMs are observed, and their intensities are identical for the two shots (Fig. 2c). On EAST, the QCM has long been observed in ELM-free H-mode plasmas(*22*), and numerical simulations show that it has the nature of a Dissipative Trapped Electron Mode (DTEM)(*26-29*). The DTEM is destabilized by the large gradient and requires high collisionality in the plasma edge(*30, 31*), which differs from the core instabilities such as the Collisionless Trapped Electron Mode (CTEM)(*32-34*). In our study, the QCM is found to be relatively low-$k$ compared with the broadband turbulence (Extended Data Fig. 3), in agreement with previous experiments and simulations.

The stationary nature of these nonlinear features in the long-pulse discharge #107124 is evident from the time evolution of the turbulence spectra (Fig. 2d), whereas in the reference discharge (Fig. 2e) the same features are also present but noticeably weaker. Cross-coherence between $k_1 = 10\ \mathrm{cm}^{-1}$ and $k_2 = 20\ \mathrm{cm}^{-1}$ shows enhanced QCM and high-frequency harmonics in #107124 relative to #107125 (Fig. 2f, g), indicating an increase in cross-scale coherent structures. The corresponding cross-phase exhibits a negative value within the high-frequency harmonic range from approximately 260 kHz to 990 kHz, with two peaks at $f_{\mathrm{1st}}$ and $f_{\mathrm{2nd}}$ (Fig. 2h,i). This negative phase provides experimental evidence consistent with forward energy transfer from low-$k$ to high-$k$. A statistical analysis of 13 H-mode shots with identical auxiliary heating power reveals an inverse relationship between broadband turbulence intensity and cross-power (Fig. 2j). Discharges with high cross-power also show a significantly negative averaged cross-coherence phase $\delta\Phi_{k1,k2}$ and elevated tungsten levels (Fig. 2k).

To quantify the nonlinear couplings more directly, we turn to bicoherence analysis. In tokamak plasmas, nonlinear couplings between disparate-scale modes can generate modulational instabilities, such as those arising from interactions between large-scale modes and small-scale turbulence(*35, 36*). Bicoherence is particularly suited to detecting such three-wave couplings. Auto-bicoherence analysis for $k_1 = 10\ \mathrm{cm}^{-1}$ at *t*=20 s in #107124 reveals nonlinear couplings between the QCM and broadband turbulence, as highlighted by the red dotted line in the bicoherence spectrum (Fig. 2l). The corresponding biphase spectrum is shown in Fig. 2m. The energy transfer function, defined as $b^2(f_1, f_2) \times \cos\beta(f_1, f_2)$, is presented in Fig. 2n, where red regions denote energy sources and blue regions denote energy sinks. The integrated energy transfer function (Fig. 2o) indicates that energy is drawn from the very spectral ranges where broadband turbulence is suppressed, and is preferentially transferred to the condensation frequencies ($f_{\mathrm{QCM}}$~30 kHz, $f_{\mathrm{1st}}$~300 kHz, $f_{\mathrm{2nd}}$~600 kHz).

Together, these findings imply that nonlinear interactions, secondary instabilities and forward energy transfer may serve as experimental indicators of confinement improvement. The suppressed broadband turbulence at $k_\theta = 10\ \mathrm{cm}^{-1}$ and $20\ \mathrm{cm}^{-1}$ aligns with the improved confinement in the core region. Although the QCM contains less than 20% of the total fluctuation energy, its nonlinear coupling to broadband turbulence reduces the latter's intensity by about 30%, implying that the weaker but coherent QCM suppresses the more transport-relevant broadband modes, and this cross-scale energy channel reshapes the fluctuation spectra in a way consistent with the improved confinement observed in discharge #107124.

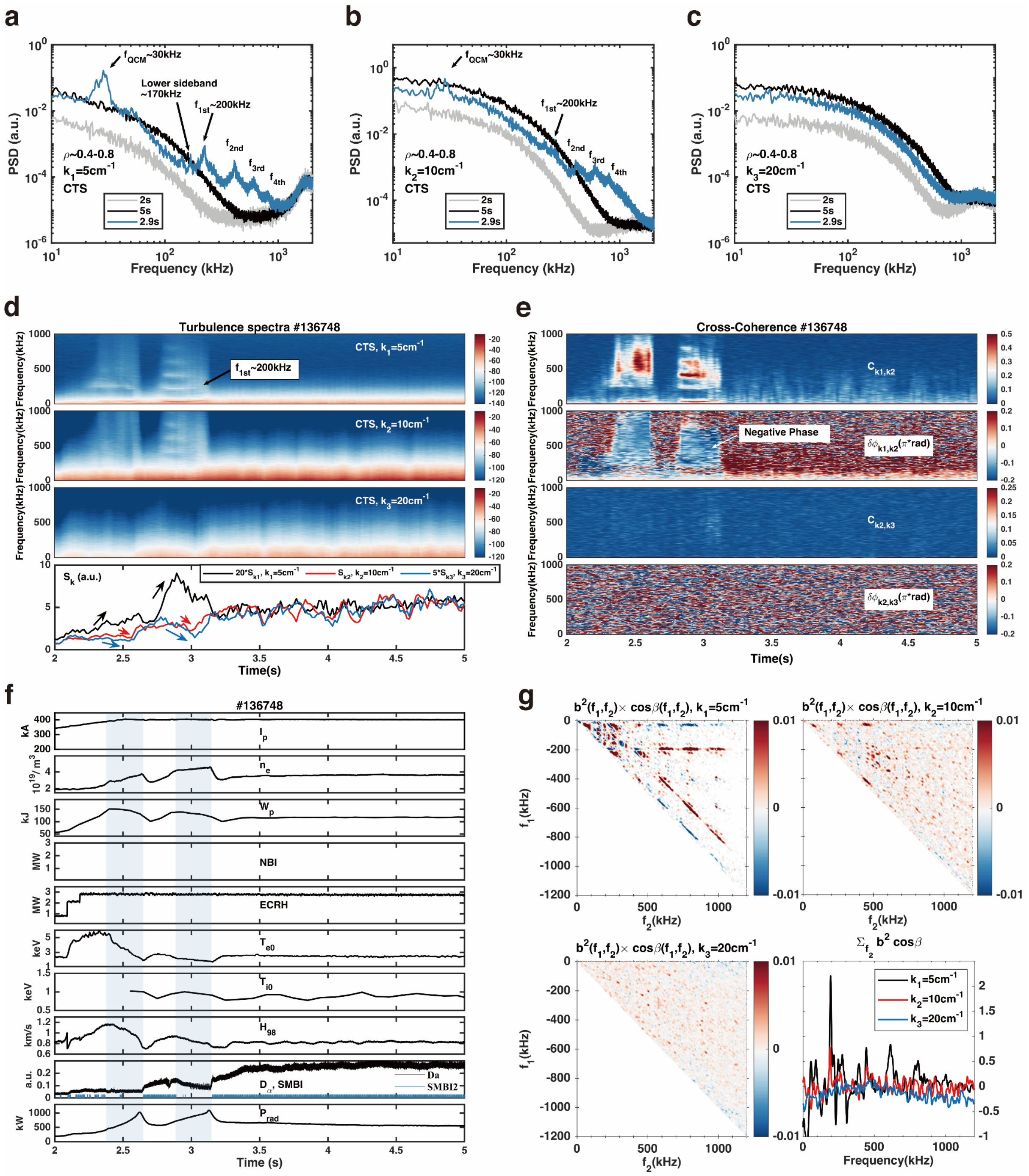

**Fig. 3 | Evidence of forward energy transfer in short-pulse discharge #136748.** **a**, Core turbulence spectra from CTS at $k_1 = 5\ \mathrm{cm}^{-1}$. **b**, Core turbulence spectra from CTS at $k_2 = 10\ \mathrm{cm}^{-1}$. **c**, Core turbulence spectra from CTS at $k_3 = 20\ \mathrm{cm}^{-1}$. **d**, From top to the bottom: time evolution of core turbulence spectra from CTS at $k_1 = 5\ \mathrm{cm}^{-1}$, $k_2 = 10\ \mathrm{cm}^{-1}$, $k_3 = 20\ \mathrm{cm}^{-1}$, and integrated turbulence intensities. **e**, From top to the bottom: time evolution of Cross-Coherence $C_{k1,k2}$, Cross-Phase $\delta\phi_{k1,k2}$,

Cross-Coherence $C_{k2,k3}$, Cross-Phase $\delta\phi_{k2,k3}$. **f**, Time evolution of main plasma parameters. **g**, Energy transfer function $b^2(f_1, f_2) \times \cos\beta(f_1, f_2)$ for $k_1 = 5$ cm$^{-1}$ (top left), $k_2 = 10$ cm$^{-1}$ (top right), $k_3 = 20$ cm$^{-1}$ (bottom left) and corresponding integrated energy transfer function $\sum_{f_2} b^2 \cos\beta$(bottom right), using signals with nonlinear interactions($t$=2.8-2.9 s).

The absence of low-$k$ core detection in the 2022 campaign, where only two CTS channels $k_\theta = 10$ cm$^{-1}$ and 20 cm$^{-1}$ were available, motivated a subsequent series of short-pulse discharges in 2024 with an additional channel at $k_\theta = 5$ cm$^{-1}$. Three CTS channels at $k_1 = 5$ cm$^{-1}$, $k_2 = 10$ cm$^{-1}$, $k_3 = 20$ cm$^{-1}$ were available, enabling simultaneous observation of ion-scale turbulence ($k_1\rho_s < 1$ in typical EAST parameters) and its coupling to electron scales. Fig. 3 presents evidence of forward energy transfer from the 2024 campaign in short-pulse discharge #136748. The turbulence spectra are compared for three representative phases: an Ohmic phase ($t$ = 2 s), an L-mode phase ($t$ = 5 s), and the H-mode phase ($t$ = 2.9 s). At the ion-scale ($k_1 = 5$ cm$^{-1}$), the characteristic signatures of forward energy transfer are clearly present at $t$ = 2.9 s, including the QCM, high-frequency harmonics ($f_{1st}$~200 kHz), and secondary instabilities. These features weaken progressively from low-$k$ to high-$k$. At $k_3 =$ 20 cm$^{-1}$, only suppressed broadband turbulence remains (Fig. 3a-c). The time evolution of the spectra and integrated intensities (Fig. 3d) reveals opposing trends between low-$k$ and high-$k$ during the formation of these patterns (two such periods occur before $t$ = 3.2 s). In the absence of these patterns (after $t$ = 3.2 s), the intensities at all three wavenumbers evolve with similar trends. The emergence of these patterns coincides with forward energy transfer, as seen in the cross-coherence and cross-phase evolution (Fig. 3e). The forward transfer signature decays with increasing wavenumber: cross-coherence and negative cross-phase between $k_2$ and $k_3$ are weaker than those between $k_1$ and $k_2$, consistent with the decay observed in the energy transfer function (Fig. 3g).

The short-pulse discharge also reveals how the forward-transfer state affects plasma control. In accordance with long-pulse discharge observations, the forward energy transfer in Fig. 3f occurs exclusively during the ELM-free H-mode phase. The radiation from high-Z impurities gradually increases during the H-mode and degrades confinement, eventually triggering an H-L transition. A key distinction from the long-pulse case is that external fueling via SMBI shuts off in the short-pulse discharge because the density rapidly exceeds the preset value, as shown by the light blue regions in Fig. 3f. By contrast, density feedback control remains active in the long-pulse discharges. This difference indicates that the improved confinement associated with forward transfer can, without active density control, lead to an uncontrolled density rise and impurity accumulation, ultimately terminating the H-mode.

The long-pulse comparison shows that the forward-transfer regime can produce a confinement bifurcation even when engineering parameters are identical. The short-pulse case further reveals that this regime can affect plasma control, as the improved confinement may drive the plasma into an uncontrolled state if density feedback is not maintained. These behaviours resemble observations associated with other confinement improvement mechanisms such as flow shear(*19, 37, 38*), negative magnetic shear(*39*), Shafranov shift(*34, 40*), or fast ion associated effects(*41, 42*), where similar bifurcations and control sensitivity have been reported. The reproducibility of the forward-transfer signatures across different heating schemes and discharge types (Extended Data Figs. 4 and 5) confirms that the regime is a robust feature of core turbulence rather than a heating-specific scenario.

**Evidence of core-edge coupling**

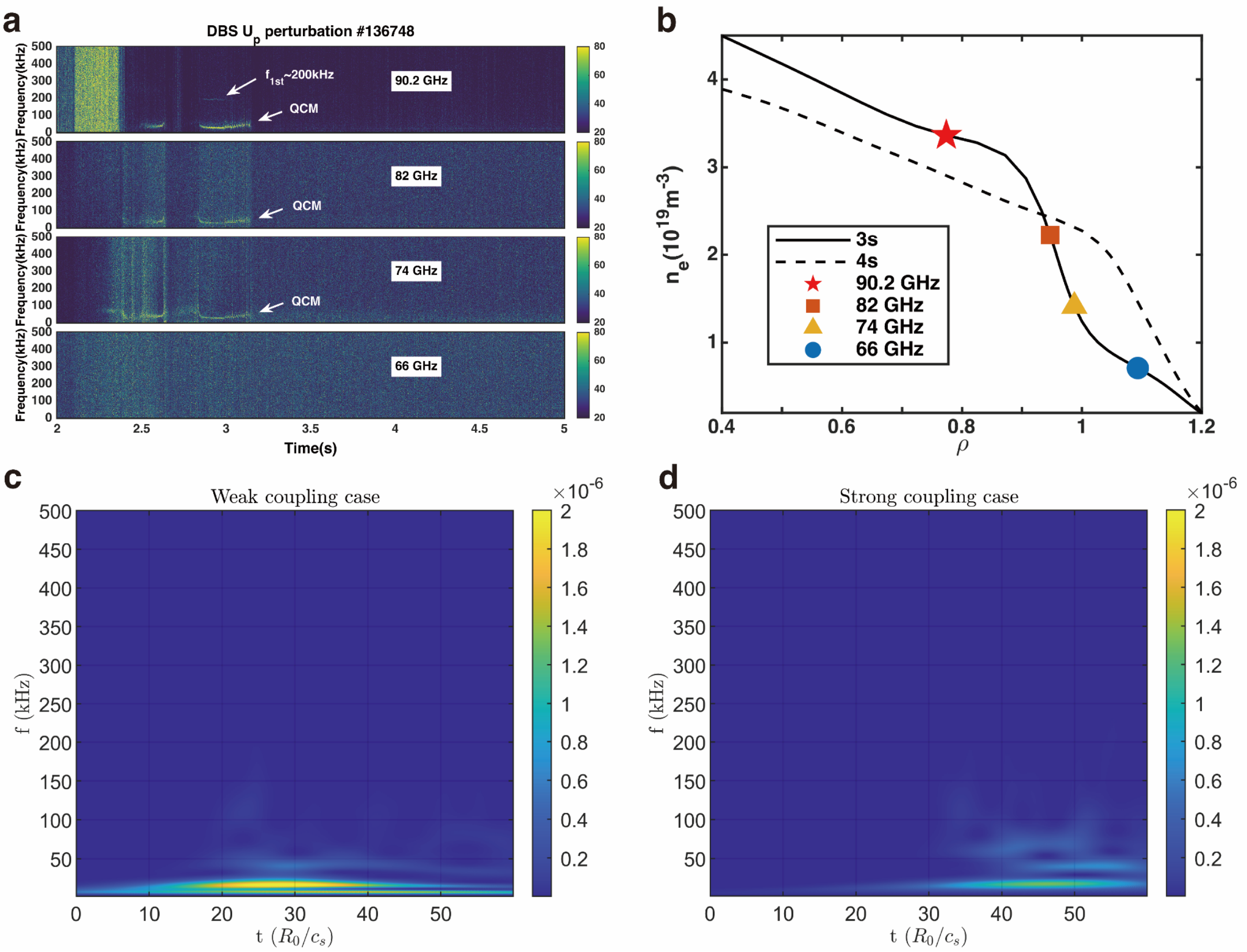

**Fig. 4 | Evidence of core-edge coupling in short-pulse discharge #136748. a**. From top to the bottom: time evolution of turbulence spectra provided by DBS with different injected electromagnetic wave frequencies, 90.2 GHz, 82 GHz, 74 GHz, 66 GHz. **b**. Locations of DBS with different injected frequencies. **c**. Wavelet spectra of turbulence in GTC nonlinear simulations, weak coupling case. d. Wavelet spectra of turbulence in GTC nonlinear simulations, strong coupling case.

While CTS provides multiscale measurements, it integrates along the scattering volume and does not resolve the precise radial location of the fluctuations. To determine whether the QCM and the high-frequency harmonics originate from the edge or the core, we turn to radially localized DBS measurements. In the EAST 2024 campaign, the upgraded DBS system was able to monitor fluctuations inside the pedestal region. Fig. 4a shows the velocity fluctuation spectra from four DBS channels with different probing frequencies, and Fig. 4b shows the density profile and the corresponding DBS measurement locations. The 66 GHz channel, located outside the last closed flux surface (LCFS, $\rho = 1$), shows no QCM. The 74 GHz and 82 GHz channels, located within the pedestal region, exhibit only a QCM. The 90.2 GHz channel, probing inside $\rho < 0.8$, shows both a QCM and the first high-frequency peak ($f_{1st}$~200 kHz for #136748). This radial dependence of QCM is in agreement with the known destabilization mechanism of DTEMs, which are driven by steep gradients in the pedestal with high collisionality. The appearance of the high-frequency peak only inside the pedestal ($\rho < 0.8$), where the QCM is also present, points to its origin in nonlinear coupling between the edge-originated QCM and core broadband turbulence, rather than being an independent edge instability.

To identify the microturbulence sources in the core and edge, we performed gyrokinetic simulations using the global gyrokinetic toroidal code (GTC)(*6*) based on the realistic equilibrium and profiles of #107124 at $t$=20 s. When the radial domain includes both core and edge regions, linear simulations show that the most unstable eigenmode near the $q$=1.5 surface exhibits typical characteristics of a CTEM: it propagates in the electron diamagnetic drift direction, is driven by trapped electrons, and is suppressed by collisions. When the simulation domain was restricted to the edge region ($\rho \sim (0.71,\ 0.94)$), a DTEM at 30 kHz became unstable. In these simulations, DTEM can be distinguished from CTEM by its sensitivity to electron collisions. CTEM is stabilized by increasing collisionality, whereas DTEM is destabilized. In nonlinear simulations, accurate modeling of core-edge coupling is limited by the lack of precise $Z_{\rm eff}$ profiles. Instead, the line-averaged $Z_{\rm eff}$ from bremsstrahlung measurements is used, which likely underestimates the edge collisionality and DTEM. Fig. 4c shows the turbulence spectra from nonlinear simulations, dominated by low-frequency components. A strong coupling case with artificially increased turbulence drive (Fig. 4d) produces higher frequency components. However, these simulations do not fully reproduce the complex spectral features observed in experiments, suggesting that insufficient experimental inputs, such as $Z_{\rm eff}$ and impurity density and temperature profiles, hinder a complete quantitative comparison.

Despite these limitations, the experimental observations and simulation results together point to a physical picture of the core-edge coupling regime, summarized in Fig. 5 and Table 2. The results suggest that the QCM, which originates in the pedestal, expands into the core and couples with broadband turbulence through nonlinear wave-wave interactions. This coupling scatters the broadband fluctuations to higher frequencies, generating cross-scale structures that dissipate through forward energy transfer, reducing the overall energy of the broadband turbulence and improving confinement. The feedback loop begins with the QCM, which originates at the edge and survives in the core, where it couples with broadband turbulence. This coupling produces secondary instabilities and drives forward energy transfer from ion scales to electron scales (steps 1 to 3). This forward transfer suppresses the broadband turbulence that drives anomalous transport, reducing thermal and particle confinement losses (steps 4 and 5). With reduced outward flux, impurities accumulate and $Z_{\rm eff}$ rises (step 6), which enhances the QCM through the collisional destabilization mechanism characteristic of DTEMs (step 7). The loop thus closes, forming a proposed self-sustaining positive feedback that simultaneously improves confinement and elevates impurity levels, resolving the apparent paradox noted in Fig. 1. Table 2 summarizes the evidence supporting each step, cross-referencing the relevant experimental figures.

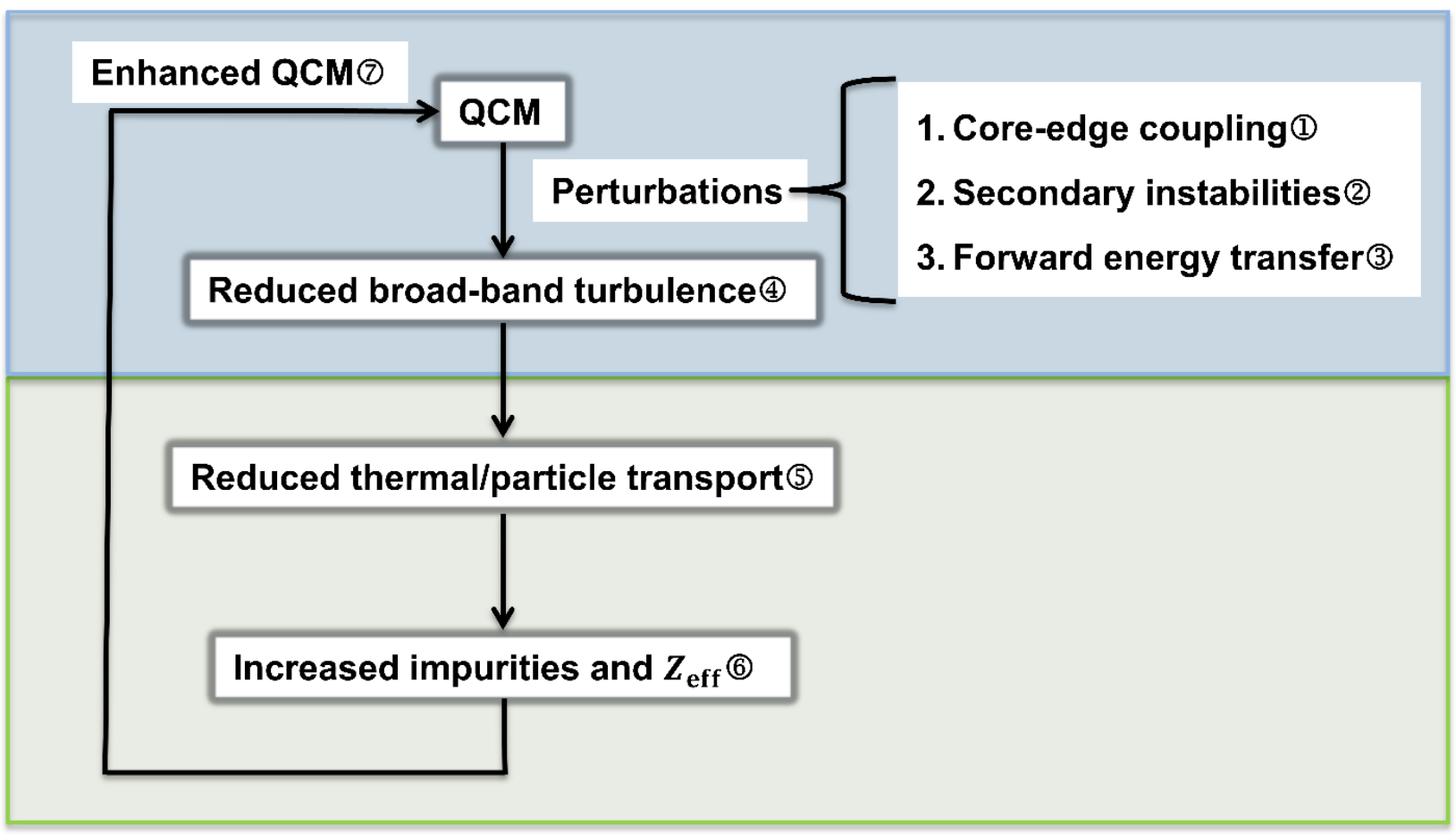

**Fig. 5 | Positive feedback loop of the core-edge coupled turbulence regime.** Schematic of the self-sustaining cycle identified in the EAST experiments. The loop starts with a QCM that originates at the edge and survives in the core, where it couples with broadband turbulence. This coupling generates secondary instabilities and drives forward energy transfer from ion scales to electron scales (steps 1 to 3). The forward transfer then suppresses the broadband turbulence that drives

anomalous transport, reducing thermal and particle transport (steps 4 and 5). With reduced outward flux, impurities accumulate and $Z_{\mathrm{eff}}$ increases (step 6), which in turn enhances the QCM (step 7) and closes the loop. See Table 2 for a detailed breakdown of the evidence for each step.

**Table 2 | Evidence chain for the core-edge coupling feedback loop.**

| Step | Description | Key evidence | Figures |
|---|---|---|---|
| ① | Core-edge coupling | QCM originates from the edge (DBS) and survives in the core (CTS), where nonlinear coupling with broadband turbulence occurs | Fig. 2a-c, Fig. 3a-c, Fig. 4a, b (DBS) |
| ② | Secondary instabilities | Sidebands satisfying $f_{\mathrm{sidebands}} = f_{\mathrm{1st}} \pm f_{\mathrm{QCM}}$ observed in CTS spectra | Fig. 2a (magnified view), Fig. 3a |
| ③ | Forward energy transfer | Negative cross-phase between low-$k$ and high-$k$ channels; energy source/sink distribution in bicoherence | Fig. 2i, Fig. 3e, Fig. 2o |
| ④ | Reduced broadband turbulence | Broadband turbulence suppressed at $k_\theta = 10\ \mathrm{cm}^{-1}$ and $20\ \mathrm{cm}^{-1}$ in improved-confinement regime | Fig. 2a,b,d, Fig. 3a–c |
| ⑤ | Reduced transport | Improved thermal and particle confinement | Fig. 1a,b,c, Extended Data Fig. 1 (lower panels) |
| ⑥ | Impurity accumulation | Elevated W-UTA signal and systematic increase of high-Z impurities (Fe, Cu, Mo, W) | Fig. 1a (W-UTA), Fig. 2k, Extended Data Fig. 2 |
| ⑦ | Enhanced QCM | Increased QCM intensity in improved-confinement regime | Fig. 2b,f,g |

## Summary and Outlook

Fifty years after CTS was first applied to multiscale turbulence measurements by Surko, Slusher, and Mazzucato, our work extends its capability to resolve cross-scale energy transfer dynamics in tokamak plasma turbulence. In the prevailing paradigm of fusion plasma turbulence, the inverse cascade and its associated zonal flows have long been considered a primary mechanism for turbulence suppression and confinement improvement. Experimental evidence of forward energy transfer, by contrast, has remained elusive, and theoretical efforts to model it have lacked clear experimental motivation. Our measurements show that quasi-two-dimensional tokamak turbulence can support a forward energy transfer from ion scales to electron scales, driven by resonant wave-wave interactions between an edge-originating QCM and core broadband turbulence. This observation establishes that forward energy transfer is not merely a theoretical possibility but an experimentally accessible pathway that actively influences confinement and plasma control. To our knowledge, this is also the first observation of modulational-instability-like spectra in tokamak core turbulence. The resolution of such fine spectral structures strengthens the evidence for nonlinear interactions and provides a new constraint for turbulence models. Across the EAST campaigns from 2021 to 2026, these signatures have been consistently observed in more than a thousand discharges, providing a robust experimental foundation for this conclusion.

The core-edge coupled regime we have identified offers a new scenario for improved confinement in the presence of high-Z impurities, a condition that is directly relevant to metal-walled devices such as ITER and CFETR. The mechanism differs from a proposed blob-void driven core-edge coupling model, which has been suggested to operate in L-mode or ELMy H-mode phases(*43*). The QCM-driven coupling in our regime appears exclusively during ELM-free H-mode, suggesting a distinct physical origin in which the edge QCM influences the core turbulence state. In long-pulse discharges, the feedback loop reaches a steady state, maintaining enhanced confinement for tens of seconds with reduced fueling demand and stable impurity radiation. In short-pulse discharges, by contrast, the same feedback leads to rapid density rise and impurity accumulation, eventually triggering an H-L back transition when the radiation exceeds the power balance limit. This contrast highlights both the promise and the challenge of this regime. It can sustain improved confinement under steady-state conditions, but it also places stringent demands on density and impurity control. Understanding and managing this feedback will be essential for exploiting its benefits in future burning plasma experiments.

Looking ahead, CTS has matured from a pioneering diagnostic in the 1970s into a powerful tool for resolving multiscale turbulence dynamics. As originally envisioned by Mazzucato(*44*), CTS based on $CO_2$ lasers offers a non-invasive, high-wavenumber-resolution approach that requires only small optical ports and is insensitive to wave refraction, making it particularly attractive for the harsh environment of burning plasmas. Our results demonstrate that CTS can reveal energy transfer channelings that were previously inaccessible. This capability will be vital for understanding and optimizing confinement in ITER and CFETR, where direct turbulence measurements are essential for validating predictive transport models and developing robust control strategies for high-performance, steady-state operation.

1. A. Alexakis, Quasi-two-dimensional turbulence. *Rev. Mod. Plasma Phys.* **7**, 31 (2023).
2. R. H. Kraichnan, Inertial Ranges in Two‐Dimensional Turbulence. *Phys. Fluids* **10**, 1417 (1967).
3. X. M. de Wit, M. Fruchart, T. Khain, F. Toschi, V. Vitelli, Pattern formation by turbulent cascades. *Nature* **627**, 515 (2024).
4. A. Fasoli, Essay: Overcoming the Obstacles to a Magnetic Fusion Power Plant. *Phys Rev Lett* **130**, 220001 (2023).
5. W. Horton, Drift waves and transport. *Rev. Mod. Phys.* **71** 735 (1999).
6. Z. Lin, T. S. Hahm, W. W. Lee, W. M. Tang, R. B. White, Turbulent Transport Reduction by Zonal Flows: Massively Parallel Simulations. *Science* **281**, 1835 (1998).
7. P. H. Diamond, S. I. Itoh, K. Itoh, T. S. Hahm, Zonal flows in plasma—a review. *Plasma Phys. Control. Fusion* **47**, R35 (2005).
8. A. Fujisawa, A review of zonal flow experiments. *Nucl. Fusion* **49**, 013001 (2009).
9. X. Si, F. De Lillo, G. Boffetta, L. Fang, Manipulating the direction of turbulent energy flux via tensor geometry in a two-dimensional flow. *Sci Adv* **11**, eadv0956 (2025).
10. H. Xia, M. G. Shats, Inverse energy cascade correlated with turbulent-structure generation in toroidal plasma. *Phys Rev Lett* **91**, 155001 (2003).
11. Z. Y. Liu, Q. G. Zong, S. Wang, X. Z. Zhou, C. Yue, Direct observations of cross-scale energy transfer driven by multiple-ion interactions in space plasmas. *Nat Commun* **16**, 11516 (2025).
12. E. Mazzucato, Small-Scale Density Fluctuations in the Adiabatic Toroidal Compressor. *Phys Rev Lett* **36**, 792 (1976).
13. C. M. Surko, R. E. Slusher, Study of the Density Fluctuations in the Adiabatic Toroidal Compressor Scattering Tokamak Using CO2 Laser. *Phys Rev Lett* **37**, 1747 (1976).

14. C. M. Surko, R. E. Slusher, Waves and Turbulence in a Tokamak Fusion Plasma. *Science* **221**, 817 (1983).
15. Q. Pratt *et al.*, Density wavenumber spectrum measurements, synthetic diagnostic development, and tests of quasilinear turbulence modeling in the core of electron-heated DIII-D H-mode plasmas. *Nucl. Fusion* **64**, 016001 (2023).
16. T. L. Rhodes *et al.*, Millimeter-wave backscatter diagnostic for the study of short scale length plasma fluctuations (invited). *Rev. Sci. Instrum* **77**, 10E922 (2006).
17. E. Mazzucato, Spectrum of Small-Scale Density-Fluctuations in Tokamaks. *Phys Rev Lett* **48**, 1828 (1982).
18. D. L. Brower, W. A. Peebles, J. N. C. Luhmann, The spectrum, spatial distribution and scaling of microturbulence in the TEXT tokamak. *Nucl. Fusion* **27**, 2055 (1987).
19. E. Mazzucato *et al.*, Turbulent Fluctuations in TFTR Configurations with Reversed Magnetic Shear. *Phys Rev Lett* **77**, 3145 (1996).
20. T. L. Rhodes *et al.*, Broad wavenumber turbulence measurements during neutral beam injection on the DIII-D Tokamak. *Nucl. Fusion* **47**, 936 (2007).
21. T. Nasu *et al.*, Electron-scale turbulence characteristics with varying electron temperature gradient in LHD. *Nucl. Fusion* **64**, 096008 (2024).
22. J. Li *et al.*, A long-pulse high-confinement plasma regime in the Experimental Advanced Superconducting Tokamak. *Nat. Phys* **9**, 817 (2013).
23. A. Loarte *et al.*, The new ITER baseline, research plan and open R&D issues. *Plasma Phys. Control. Fusion* **67**, 065023 (2025).
24. E. J. Doyle *et al.*, Chapter 2: Plasma confinement and transport. *Nucl. Fusion* **47**, S18 (2007).
25. L. Zhang *et al.*, A fast-time-response extreme ultraviolet spectrometer for measurement of impurity line emissions in the Experimental Advanced Superconducting Tokamak. *Rev. Sci. Instrum* **86**, 123509 (2015).
26. H. Q. Wang *et al.*, New edge coherent mode providing continuous transport in long-pulse H-mode plasmas. *Phys Rev Lett* **112**, 185004 (2014).
27. Y. Ye *et al.*, A stationary long-pulse ELM-absent H-mode regime in EAST. *Nucl. Fusion* **57**, 086041 (2017).
28. C. Zhao, T. Zhang, Y. Xiao, Gyrokinetic simulation of dissipative trapped electron mode in tokamak edge. *Phys. Plasmas* **24**, 052509 (2017).
29. C. Zhao, Y. Xiao, Nonlinear gyrokinetic particle simulation of dissipative trapped electron mode. *Phys. Plasmas* **25**, 102303 (2018).
30. W. X. Wang *et al.*, Distinct turbulence sources and confinement features in the spherical tokamak plasma regime. *Nucl. Fusion* **55**, 122001 (2015).
31. W. X. Wang *et al.*, Identification of new turbulence contributions to plasma transport and confinement in spherical tokamak regime. *Phys. Plasmas* **22**, 102509 (2015).
32. Y. Xiao, Z. Lin, Turbulent transport of trapped-electron modes in collisionless plasmas. *Phys Rev Lett* **103**, 085004 (2009).
33. H. Chen, L. Chen, On the cascading of collisionless trapped-electron mode turbulence in tokamak plasmas. *Nucl. Fusion* **59**, 074003 (2019).
34. F. Chen *et al.*, Strong electron temperature stiffness induced by broad-band turbulence in high poloidal beta plasmas on EAST. *Nucl. Fusion* **65**, 046014 (2025).
35. L. Bardoczi *et al.*, Modulation of Core Turbulent Density Fluctuations by Large-Scale Neoclassical Tearing Mode Islands in the DIII-D Tokamak. *Phys Rev Lett* **116**, 215001 (2016).
36. G. D. Conway, A. I. Smolyakov, T. Ido, Geodesic acoustic modes in magnetic confinement devices. *Nucl. Fusion* **62**, 013001 (2022).
37. D. R. Smith *et al.*, Observations of reduced electron Gyroscale fluctuations in national spherical torus experiment H-mode plasmas with large ExB flow shear. *Phys Rev Lett* **102**, 225005 (2009).
38. P. Mantica *et al.*, A key to improved ion core confinement in the JET tokamak: ion stiffness mitigation due to combined plasma rotation and low magnetic shear. *Phys Rev Lett* **107**, 135004 (2011).
39. H. Y. Yuh *et al.*, Suppression of electron temperature gradient turbulence via negative magnetic shear in NSTX. *Phys Rev Lett* **106**, 055003 (2011).
40. J. McClenaghan *et al.*, Shafranov shift bifurcation of turbulent transport in the high βp scenario on DIII-D. *Nucl. Fusion* **59**, 124002 (2019).
41. H. Han *et al.*, A sustained high-temperature fusion plasma regime facilitated by fast ions. *Nature* **609**, 269 (2022).
42. Y.-S. Na *et al.*, How fast ions mitigate turbulence and enhance confinement in tokamak fusion plasmas. *Nat. Rev. Phys.* **7**, 190 (2025).

43. M. Cao, P. H. Diamond, Physics of Edge-Core Coupling by Inward Turbulence Propagation. *Phys Rev Lett* **134**, 235101 (2025).
44. E. Mazzucato, Detection of short-scale turbulence in the next generation of tokamak burning plasma experiments. *Plasma Phys. Control. Fusion* **48**, 1749 (2006).
45. A. Hasegawa, K. Mima, Pseudo-three-dimensional turbulence in magnetized nonuniform plasma. *Phys. Fluids* **21**, 87 (1978).
46. P. H. Diamond, A. Hasegawa, K. Mima, Vorticity dynamics, drift wave turbulence, and zonal flows: a look back and a look ahead. *Plasma Phys. Control. Fusion* **53**, 124001 (2011).
47. Y. Ren *et al.*, Density gradient stabilization of electron temperature gradient driven turbulence in a spherical tokamak. *Phys Rev Lett* **106**, 165005 (2011).
48. P. Li *et al.*, Study of turbulence modulation and core density peaking with CO2 laser collective scattering diagnostics in the EAST tokamak. *Nucl. Fusion* **60**, 066001 (2020).
49. W. Lee *et al.*, Development of a collective scattering system and its application to the measurement of multiscale fluctuations in KSTAR plasmas. *Plasma Phys. Control. Fusion* **63**, 035003 (2021).
50. D. C. Speirs *et al.*, Simulation and analysis of a high-k electron-scale turbulence diagnostic for MAST-U. *Nucl. Fusion* **65**, 046019 (2025).
51. T. Tokuzawa *et al.*, Cross-scale nonlinear interaction and bifurcation in multi-scale turbulence of high-temperature plasmas. *Commun Phys* **8**, 394 (2025).
52. S. Maeyama *et al.*, Multi-scale turbulence simulation suggesting improvement of electron heated plasma confinement. *Nat Commun* **13**, 3166 (2022).
53. A. W. Leonard, Edge-localized-modes in tokamaks. *Phys. Plasmas* **21**, 090501 (2014).
54. H. Arnichand *et al.*, Quasi-coherent modes and electron-driven turbulence. *Nucl. Fusion* **54**, 123017 (2014).

**Acknowledgements**: The authors would like to thank W.X. Wang for fruitful discussions and Yuehao Ma and Hanyang Lv for their assistance with verifying the simulation results. This work was supported by the National MCF Energy R&D Program (2025YFE03200200, 2022YFE03050000), the National Natural Science Foundation of China under Contract Nos. 12605372, 12605373, 12305254, the Strategic Priority Research Program of Chinese Academy of Sciences under Grant No. XDB0790101, XDB0790202.

* email: fei.chen@ipp.ac.cn, lipan@ipp.ac.cn, pfliu@iphy.ac.cn, j_li@ipp.ac.cn

## Methods

**EAST tokamak**: EAST is a fully superconducting tokamak with a major radius $R_0$ = 1.85 m and a minor radius a = 0.45 m(*22*). The experiments are conducted in double null (DN) cross section, and the auxiliary heating power sources include Lower Hybrid Wave (LHW), Ion Cyclotron Range of Frequencies (ICRF), and Electron Cyclotron Resonance Heating (ECRH). The toroidal magnetic field strength is 2.48 T.

**Collective Thomson scattering (CTS)**: CTS using electromagnetic waves (lasers or microwaves) as probes offers a non-invasive method for measuring multiscale turbulence in such high-temperature plasmas. In their landmark studies, Surko and Slusher used laser scattering(*13*), whereas Mazzucato used microwave scattering(*12*), to measure multiscale turbulence in tokamaks. The observed spectra, which were significantly broader than weak-turbulence predictions, motivated the development of strong-turbulence models that incorporate inverse energy transfer(*45, 46*). Since then, collective scattering has been widely deployed across devices to study core turbulence, including PLT(*14, 17*), TEXT(*18*), TFTR(*19*), DIII-D(*20*), NSTX(*37, 39, 47*), and more recently on EAST(*48*), KSTAR(*49*), MAST-U(*50*), and LHD(*21, 51*).

**EAST campaigns**: Two main series of discharges are reported in this article: long-pulse discharges in EAST campaign 2022 and short-pulse discharges in EAST campaign 2024. In EAST campaign 2022, only two wavenumbers, $k_\theta = 10$ cm$^{-1}$ and $k_\theta = 20$ cm$^{-1}$, were available for CTS, and the highest injected microwave frequency of Doppler Back Scattering (DBS) was 75 GHz, so DBS mainly detected turbulence in the plasma edge. A low-$k$ channel at $k_\theta = 5$ cm$^{-1}$ became available for CTS after 2023, and DBS channels with higher microwave frequencies were able to probe core turbulence.

**Cross-coherence analysis**: Cross-coherence is defined as $C_{x,y}(f) = |P_{x,y}(f)|^2/|P_{x,x}(f)||P_{y,y}(f)|$, which indicates how well signal x corresponds to signal y at each frequency. The $P_{x,y}(f)$ is cross power spectral density, while $P_{x,x}(f)$ and $P_{y,y}(f)$ are auto-power spectral densities. The corresponding cross-phase is $\delta\phi_{x,y} = tan^{-1}\frac{Im(P_{x,y}(f))}{Re(P_{x,y}(f))}$. For two CTS channels ($x = k_1, y = k_2$), the cross-power is defined as $S_{\text{cross}-k1,k2} = \sum_f C_{k1,k2}(f)$ and averaged cross-phase is $\delta\Phi_{k1,k2} = \frac{\sum_f C_{k1,k2}(f)\delta\phi_{k1,k2}(f)}{\sum_f C_{k1,k2}(f)}$.

**Bicoherence and energy transfer function analysis**: Auto-bicoherence analysis is used to quantify nonlinear three-wave coupling. The auto-bicoherence spectrum is defined as $b^2(f_1, f_2) = \frac{|B(f_1,f_2)|^2}{\langle P(f_1)\rangle\langle P(f_2)\rangle\langle P(f_1+f_2)\rangle}$, $B(f_1, f_2) = \langle P(f_1)P(f_2)P^*(f_1 + f_2)\rangle$, where P(f) is the Fourier transform of the turbulent signal and the angle brackets denote an ensemble average. The associated biphase is $\beta(f_1, f_2) = \arctan\frac{\mathrm{Im}[B(f_1,f_2)]}{\mathrm{Re}[B(f_1,f_2)]}$. The energy transfer function is then defined as $T(f_1, f_2) = b^2(f_1, f_2) \cdot \cos[\beta(f_1, f_2)]$, where $T > 0$ indicates an energy source (transfer into mode $f_1 + f_2$) and $T < 0$ indicates an energy sink (transfer out of the interacting modes). The integrated energy transfer is computed as $I_T(f) = \sum_{f_2} T(f, f_2)$.

## Glossary

**Multiscale turbulence**: Magnetic fusion plasma is a multi-scale problem involving wide temporal and spatial scales, from the electron gyroradius (<0.1 mm) to the ion gyroradius (<1 cm). In the context of multiscale tokamak turbulence, 'ion-scale' refers to fluctuations with perpendicular wavenumbers comparable to the ion sound gyroradius ($k_\perp\rho_s < 1$, $\rho_s$ is the ion sound Larmor radius), whereas 'electron-scale' denotes modes smaller than ion-scale ($k_\perp\rho_s > 1$). The cross-scale interactions between electron-scale and ion-scale turbulence can affect turbulent transport of particles and energy(*52*).

**Edge-localized mode (ELM)**: A magneto-hydrodynamic instability that occurs in the H-mode pedestal region of tokamak plasmas(*53*). When the pressure and current gradients in the edge transport barrier grow beyond the MHD stability limit, the ELM instability grows explosively, rapidly transporting energy and particles onto open field lines and material surfaces. ELMs provide additional particle and impurity transport through the H-mode transport barrier, enabling steady operation, but the resulting heat flux transients pose a risk to plasma facing components in future burning plasma devices.

**Quasi-coherent mode (QCM) and broadband turbulence**: In tokamak turbulence, broadband fluctuations are stochastic, large-amplitude fluctuations that span a wide frequency range ($\Delta f \approx$ hundreds of kHz) and are considered the signature of fully developed turbulence. Coherent modes, by contrast, are narrow-band fluctuations with a well-defined oscillation frequency ($\Delta f \approx$ few kHz). A quasi-coherent mode shares characteristics of both. It oscillates around a given frequency but exhibits significant frequency broadening ($\Delta f \approx$ tens of kHz), indicating coupling to the turbulent background. See reference(*54*) for the details. In this work, the QCM refers to such a mode observed in EAST ELM-free H-mode plasmas(*26*).

**Collisionless trapped electron mode (CTEM) and dissipative trapped electron mode (DTEM)**: Trapped electron modes (TEMs) are electrostatic instabilities driven by trapped electrons and are a major source of anomalous transport in tokamaks. The collisionless trapped electron mode (CTEM) is destabilized through the trapped electron magnetic precessional drift resonance with the mode and is widely believed to be important in the core of conventional tokamak plasmas with low collisionality. The dissipative trapped electron mode (DTEM), by contrast, arises from the scattering of trapped electrons into passing electrons due to collisions and is usually considered negligible in the core region, where low collisionality naturally stabilizes the mode. In this work, the edge QCM is identified as a DTEM through gyrokinetic simulations.

**Modulational instability**: A nonlinear instability in which a test large-scale, slow perturbation on an ensemble of waves grows by triggering a flow of energy that reinforces the original perturbation. It arises from the interplay between nonlinear and dispersive effects and can be understood as a four-wave interaction. Classic examples include the self-focusing of a beam and the amplification of shear flows by drift-wave turbulence. In this work, the observed secondary sidebands and energy transfer function bear signatures of modulational instability, with the high-frequency peak acting as the pump wave and the QCM serving as the modulating wave.

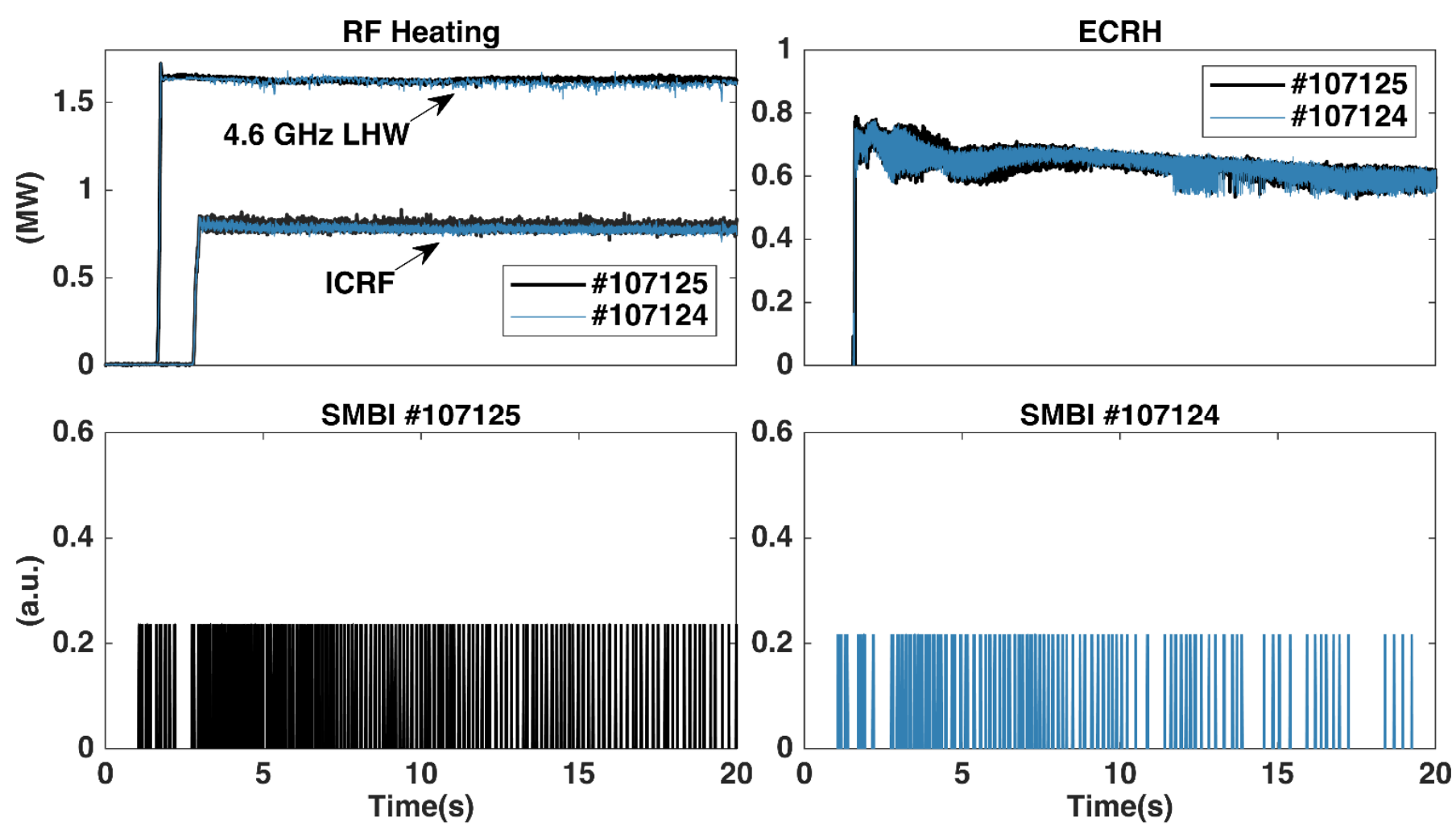

**Extended Data Fig. 1 | Identical heating scenarios for #107124/#107125 (upper panels) and reduced fuel rate for #107124 (bottom panels).**

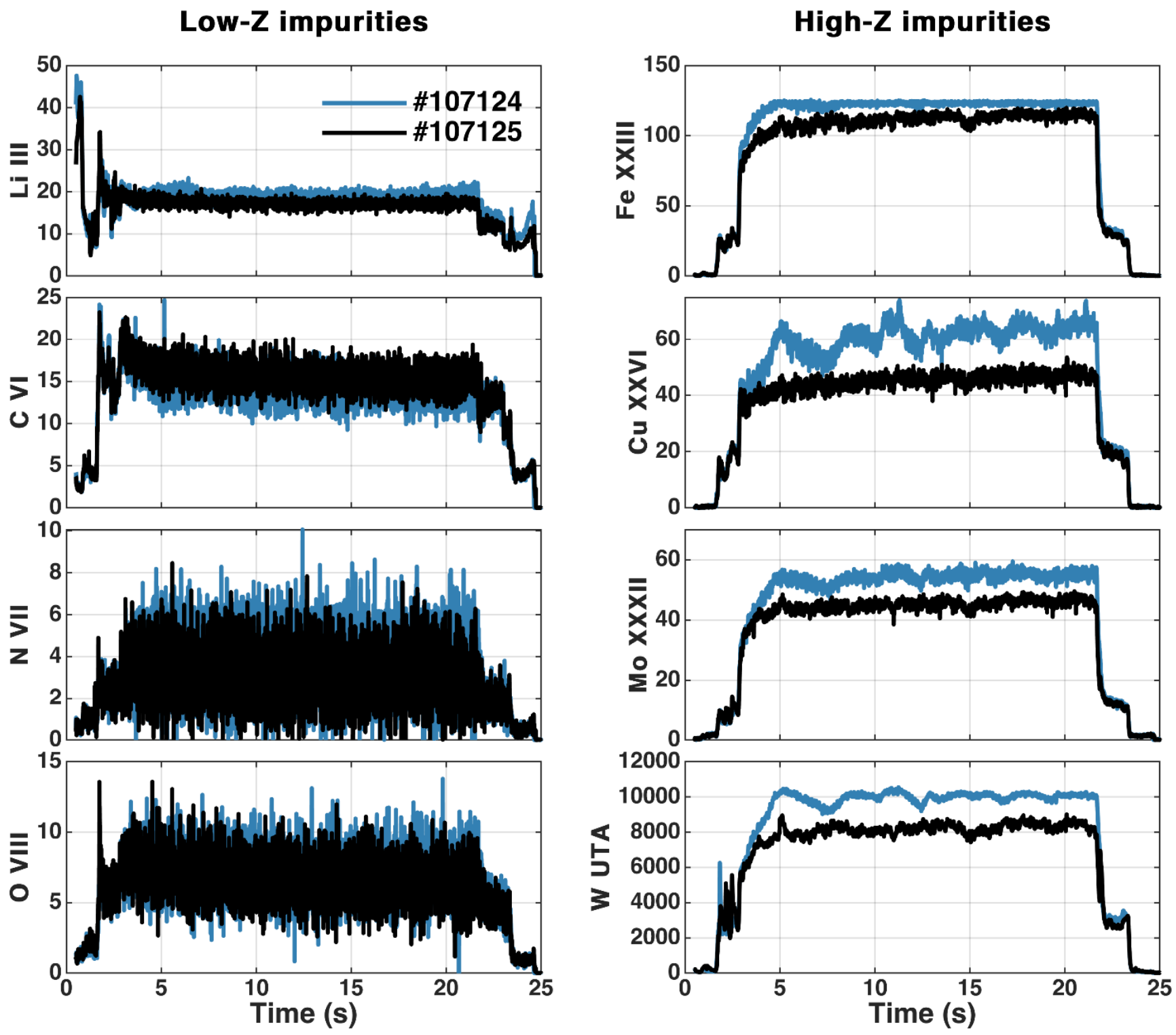

**Extended Data Fig. 2 | Identical low-Z impurities for #107124/#107125 (left panels, from top to the bottom): lithium, carbon, nitrogen, oxygen. Enhanced high-Z impurities for #107124 (right panels, from top to the bottom): iron, copper, molybdenum, tungsten.**

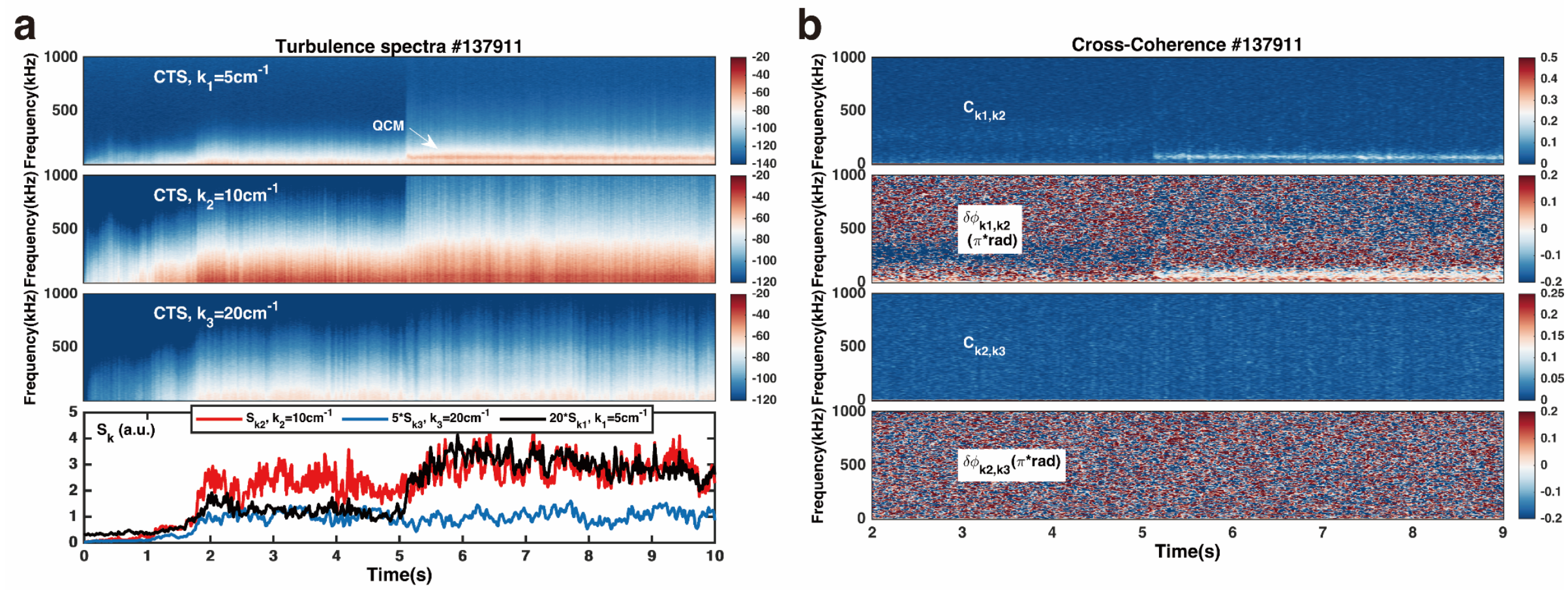

**Extended Data Fig. 3 | Low-*k* nature of QCM in discharge #137911 (without observable nonlinear interactions). a**. From top to the bottom: time evolution of core turbulence spectra provided by CTS at $k_1 = 5\ \mathrm{cm}^{-1}$, $k_2 = 10\ \mathrm{cm}^{-1}$, $k_3 = 20\ \mathrm{cm}^{-1}$, and integrated turbulence intensities. It is shown that QCM appears at about *t*=5s and increases the integrated turbulence intensities at low-*k* and intermediate-*k*, but high-*k* is nearly not affected. **b**. From top to the bottom: time evolution of Cross-Coherence $C_{k1,k2}$, Cross-Phase $\delta\phi_{k1,k2}$, Cross-Coherence $C_{k2,k3}$, Cross-Phase $\delta\phi_{k2,k3}$.

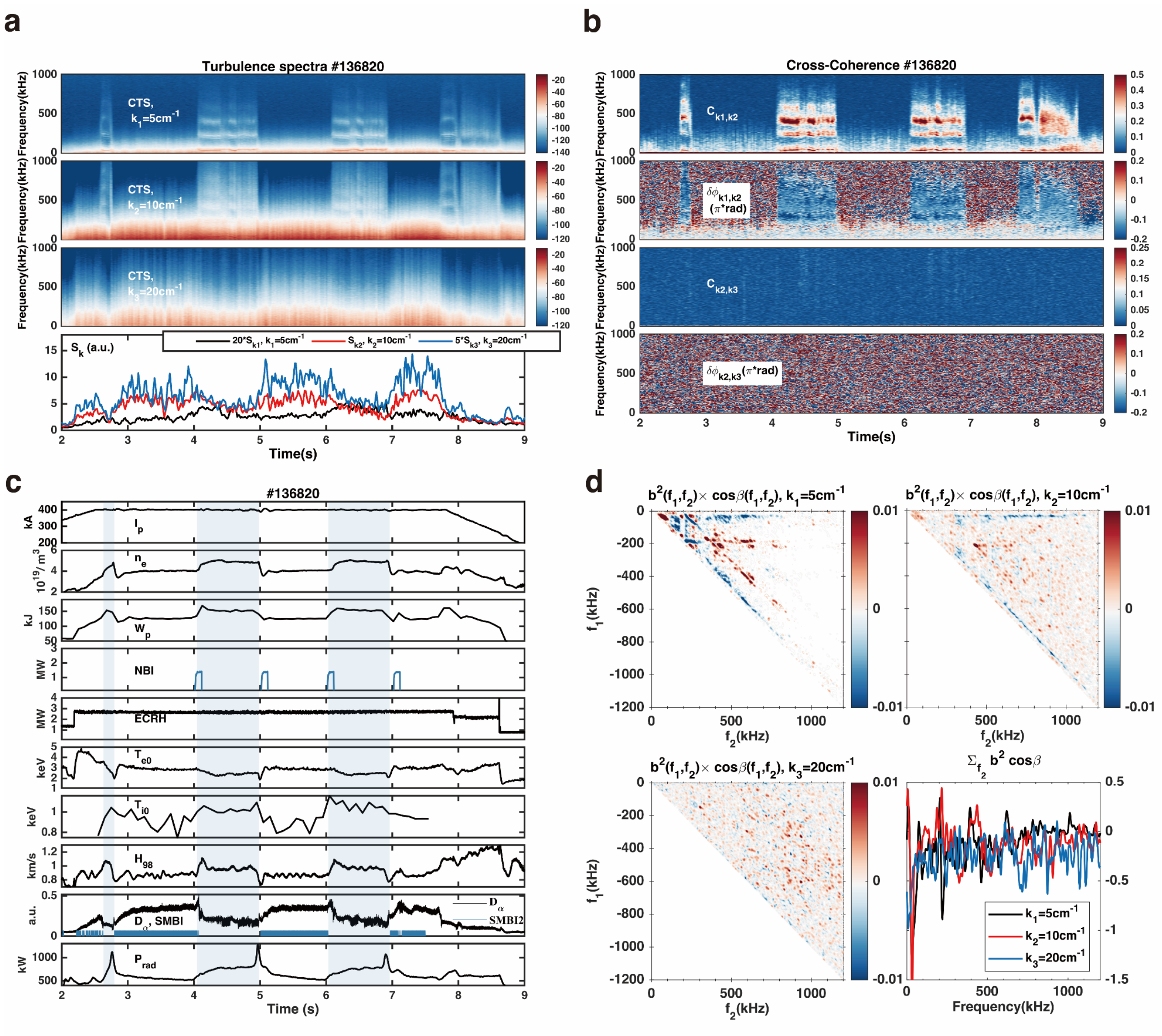

**Extended Data Fig. 4 | Evidence of forward energy transfer in discharge #136820, where the L-H transitions are induced by NBI.** **a**. From top to the bottom: time evolution of core turbulence spectra provided by CTS at $k_1 = 5\ \mathrm{cm}^{-1}$, $k_2 = 10\ \mathrm{cm}^{-1}$, $k_3 = 20\ \mathrm{cm}^{-1}$, and integrated turbulence intensities. **b**. From top to the bottom: time evolution of Cross-Coherence $C_{k1,k2}$, Cross-Phase $\delta\phi_{k1,k2}$, Cross-Coherence $C_{k2,k3}$, Cross-Phase $\delta\phi_{k2,k3}$. **c**. Time evolution of main plasma parameters. **d**. Energy transfer function $b^2(f_1, f_2) \times \cos\beta(f_1, f_2)$ for $k_1 = 5\ \mathrm{cm}^{-1}$ (top left), $k_2 = 10\ \mathrm{cm}^{-1}$ (top right), $k_3 = 20\ \mathrm{cm}^{-1}$ (bottom left) and corresponding integrated energy transfer function $\sum_{f_2} b^2 \cos\beta$(bottom right), using signals with nonlinear interactions($t$=4.2-4.3 s).

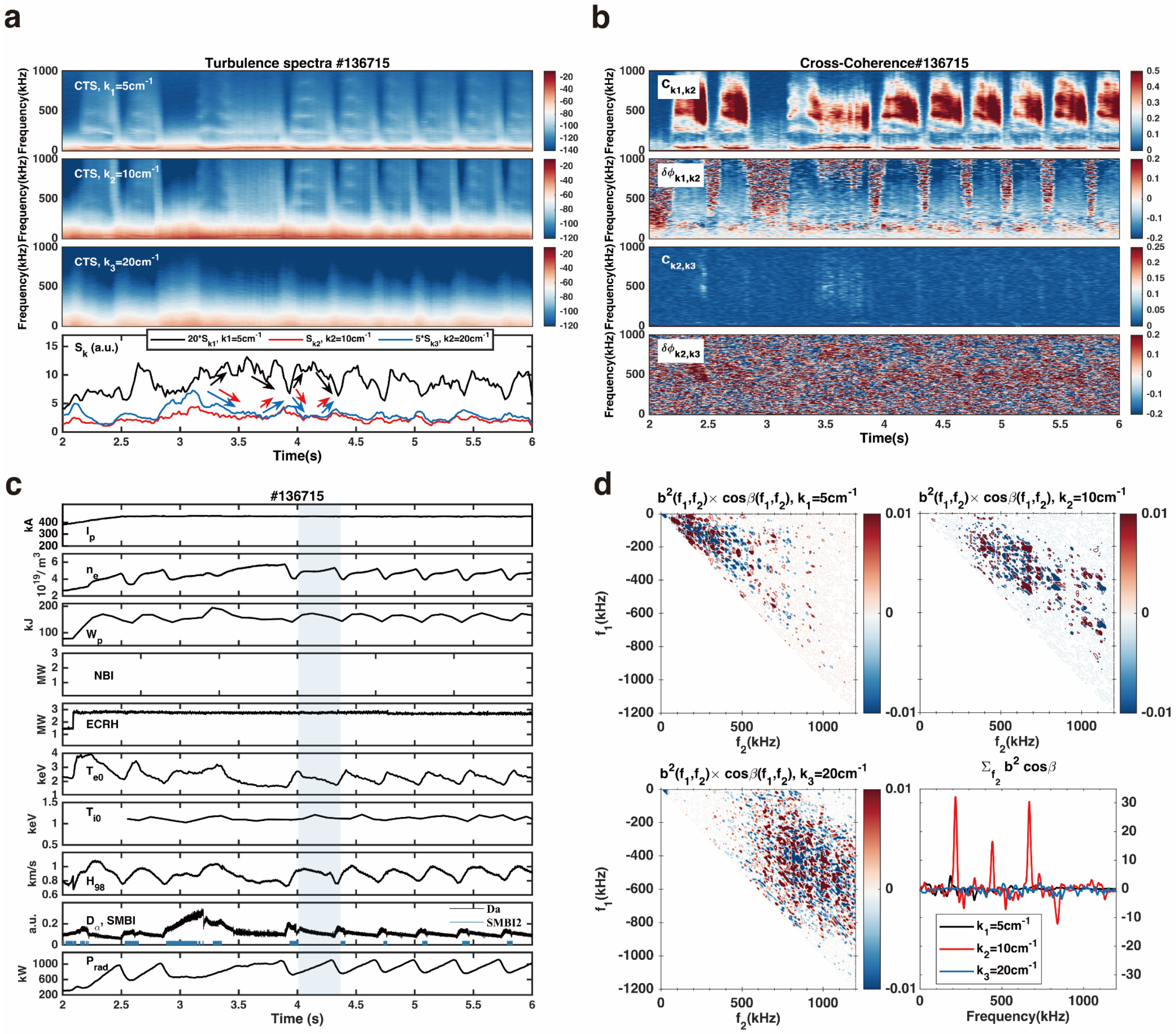

**Extended Data Fig. 5 | Evidence of forward energy transfer in discharge #136715, where the L-H transitions lead to spontaneous oscillations.** **a**. From top to the bottom: time evolution of core turbulence spectra provided by CTS at $k_1 = 5\ \mathrm{cm}^{-1}$, $k_2 = 10\ \mathrm{cm}^{-1}$, $k_3 = 20\ \mathrm{cm}^{-1}$, and integrated turbulence intensities. **b**. From top to the bottom: time evolution of Cross-Coherence $C_{k1,k2}$, Cross-Phase $\delta\phi_{k1,k2}$, Cross-Coherence $C_{k2,k3}$, Cross-Phase $\delta\phi_{k2,k3}$. **c.** Time evolution of main plasma parameters. **d**. Energy transfer function $b^2(f_1, f_2) \times \cos\beta(f_1, f_2)$ for $k_1 = 5\ \mathrm{cm}^{-1}$ (top left), $k_2 = 10\ \mathrm{cm}^{-1}$ (top right), $k_3 = 20\ \mathrm{cm}^{-1}$ (bottom left) and corresponding integrated energy transfer function $\sum_{f_2} b^2 \cos\beta$(bottom right), using signals with nonlinear interactions($t$=3.5-3.6 s).